\documentclass[aps,pre,twocolumn,superscriptaddress]{revtex4}

\usepackage{amssymb} 
\usepackage{amsmath} 
\usepackage{graphicx}
\usepackage{float}

\newcommand*{\m}{\text{m}} 
\newcommand*{\intpl}{\int_{x-\frac{1}{2}+\xi}^{x+\frac{1}{2}}d\zeta\;}
\newcommand*{\intmi}{\int_{x-\frac{1}{2}}^{x+\frac{1}{2}+\xi}d\zeta\;}

\begin{document}

\title{Self-organization and Mechanical Properties of Active Filament Bundles}

\author{Karsten Kruse}  \email[]{karsten@mpipks-dresden.mpg.de}
\affiliation{Max-Planck-Institut f\"ur Physik komplexer Systeme, N\"othnitzer
Str.~38, 01187 Dresden, Germany}  \affiliation{Institut Curie, Physicochimie,
UMR CNRS/IC 168, 26 rue d'Ulm, 75248 Paris Cedex 05, France}
\affiliation{Max-Planck-Institut f\"ur Str\"omungsforschung, Bunsenstr.~10,
37073 G\"ottingen, Germany}  \author{Frank J\"ulicher}
\email[]{julicher@mpipks-dresden.mpg.de} \affiliation{Max-Planck-Institut
f\"ur Physik komplexer Systeme, N\"othnitzer Str.~38, 01187 Dresden, Germany}
\affiliation{Institut Curie, Physicochimie, UMR CNRS/IC 168, 26 rue d'Ulm,
75248 Paris Cedex 05, France}

\date{\today}

\begin{abstract}
A general framework for the description of active bundles of polar filaments
is presented. The activity of the bundle results from mobile cross-links, that
induce  relative displacements between the aligned filaments.  Our generic
description is based on momentum conservation within the bundle. By specifying
the internal forces, a simple minimal model for the bundle dynamics can be
derived, capturing a rich variety of dynamic behaviors.  In particular,
contracted states as well as solitary and oscillatory waves appear through
dynamic instabilities.  We present the full bifurcation diagram of this model
and study the effects of a dynamic motor distribution on the bundle dynamics.
Furthermore, we discuss the mechanical properties of the bundle in the
presence of externally applied forces. Our description is motivated by dynamic
phenomena in the cytoskeleton and could apply to in vitro experiments as well
as to stress-fibers and to self-organization phenomena during cell-locomotion.

\end{abstract}


\maketitle

\section{Introduction}

The cytoskeleton of eucaryotic cells is a complex three dimensional network of
protein filaments, most prominently actin filaments and
microtubules~\cite{ajlrrw02,bray01}. Its elastic and viscous properties are
essentially defining the mechanical or material properties of living
cells. This network resembles in many aspects a polymer  solution or a gel.
The main difference from usual polymer materials is its intrinsic activity. In
fact, the cytoskeleton is constantly remodeled through the polymerization and
depolymerization of filaments, as well as through the formation and breakup of
cross-links. In addition, the cross-links may be active, leading to further
dynamics.  Active or mobile cross-links are provided for example by molecular
motors which are specialized enzymes which transduce the chemical energy of a
fuel to motion along filaments~\cite{ajlrrw02,bray01,kv93,jap97}.  All these
activities are regulated by the cell which is thus able to direct
intracellular transport, to separate its chromosomes and to cleave during cell
division, to exert forces on the environment, or to move on a substrate.

The study of active polymer systems requires completely new tools and
techniques as compared to the well developed analysis of equilibrium
properties, that relies on powerful concepts of equilibrium statistical
physics. Indeed, such systems are intrinsically far from equilibrium, and the
dynamics at equilibrium, that is usually studied in polymer
physics~\cite{amypl96,mj97,morse98a}, is not sufficient for the description of
active systems \cite{lego02}. On the contrary, experimental studies of the
cytoskeleton under simplified conditions have revealed its ability to
self-organize. Namely, the contraction of filament bundles~\cite{takiguchi91},
the formation of asters and vortices~\cite{umamk91,nsml97,sewynsl98,ns01}, as
well as the formation of networks~\cite{snlk01} were found in vitro. Using a
cell extract, even the formation of bipolar spindles without microtubule
organizing centers has been seen~\cite{hk96}. Cell fragments containing only
the actin cytoskeleton, but neither the nucleus nor microtubules, can
propagate on a substrate~\cite{es84}, where the locomoting state coexists with
a stationary spherically symmetric state~\cite{vsb99}. In a mixture of actin
filaments and myosin molecular motors, active reptation in a polymer solution
has been observed~\cite{hdssk02}. Let us finally mention, that experiments
probing mechanical properties of living cells have revealed active responses
of the cytoskeleton to external forces, see, e.g., Ref.~\cite{to99}.

First steps towards a theoretical understanding of active polymer systems have
mostly aimed at describing pattern-formation. In one-dimensional filament
bundles, polarity sorting~\cite{ns96}, contraction~\cite{sn96,kj00}, and
propagating waves~\cite{kcj01} have been observed.  Self-organization has also
been seen to induce bending waves and complex motion in
axonemes~\cite{cjp99,cj00}. In higher dimensions, the effects of active
cross-links on the formation of orientation patterns in systems of spatially
fixed  filaments have been studied~\cite{blj00,lk01} and the generation of
filament currents by active cross-links has been
discussed~\cite{live02}. Furthermore, the viscoelastic response of solutions
of semi-flexible polymers and active centers has been studied~\cite{lma01}.

Active filament bundles provide very simple examples of active filament
networks and can be discussed by a one-dimensional description. Note, however,
that in addition   to their simplicity, such filament bundles actually occur
in animal cells. They are, for example, part of stress fibers which generate
contractile forces, and of the contractile ring in  dividing
cells~\cite{bray01}. In vitro, the contraction of  actin bundles in the
presence of myosin motors has been observed~\cite{takiguchi91}.

Motivated by the dynamics of the cytoskeleton, we develop a general framework
based on momentum conservation, to describe the physics of bundles of  aligned
filaments in the presence of active cross-links. Since  actin filaments and
microtubules are polar as they have two structurally different ends, we
consider polar filaments. The cross-links are mobile and considered to be
formed by small aggregates of molecular motors of one type.  We  discuss
simple scenarios in order to study dynamic phenomena and mechanical properties
of such systems. A minimal model which has been introduced in earlier
publications~\cite{kj00,kcj01}  can be derived in our general framework using
approximations and simplifications. This model already exhibits many of the
phenomena that occur in such systems. However, it neglects changes in the
distribution of motors due to the dynamics of the system.  This dynamics of
the motor distribution can be taken into account within our general
framework. Furthermore, we discuss the mechanical properties of active bundles
in the presence and absence of external forces.

The outline of our manuscript is as follows. In section~\ref{sec:grundlagen},
we introduce  the general description for active filament bundles based on
momentum balance. Using this formalism we derive dynamic equations for the
system. In Section~\ref{sec:model1}, we show how the minimal model can be
derived from the general equations. We review its properties and extend
previous work  towards a discussion of the full bifurcation diagram.  In
Section~\ref{sec:model2} we study the effects of the dynamics of the motor
distribution on the filament dynamics.  The active mechanical properties of a
filament bundle can be derived systematically in the framework introduced in
section~\ref{sec:grundlagen}.  We discuss the bundle mechanics  in
Section~\ref{sec:spannung} and study tense states balanced by external forces
applied at the ends.  The paper concludes with a discussion of our results in
Section~\ref{sec:diskussion} which relates our theoretical framework to
experimental situations.  The appendices contains a detailed analysis of
bifurcations in the minimal model as well as the effects of filament adhesion
to a substrate.

\section{\label{sec:grundlagen} Phenomenological description of active
filament bundles}

We introduce a general description for the dynamics of active filament
bundles. The bundle is described in one dimension using density profiles of
filaments and motors. The dynamics of these densities is governed by currents
which are generated by interactions between filaments and motors. Dynamic
equations can be derived on the basis of momentum balance. This general
procedure can be carried out most conveniently using simplifying assumptions.
In particular, we assume  a low motor density or low duty ratio of motors such
that interactions between filament pairs dominate, a local friction of
filaments with the environment and we neglect the possibility of passive
cross-linkers. Furthermore, we assume that filament lengths remain fixed,
i.e. we neglect polymerization and depolymerization of filaments, and we
assume that filaments cannot change their orientation.  However, many of the
qualitative behaviors displayed by the resulting equations are more general
and are also found in purely phenomenological descriptions which  are not
based on these simplifying assumptions \cite{kjup}.  The dynamic equations we
discuss here represent a mean field theory of filament bundles where
fluctuations do not appear explicitly but  give rise to diffusive terms.

\subsection{Densities of filaments and motors}

The bundle is characterized by the number densities of filaments and of motor
complexes projected on the bundle axis, which leads to an effective
one-dimensional description.  Since filament bending and entanglements can be
ignored in the bundle, we describe filaments as rigid rods.  Filaments are
aligned along the $x$-axis and we distinguish the two sub-populations of
filaments with their plus-end pointing into the positive and negative
$x$-direction, respectively. These populations are described by the  densities
$c^+$ and $c^-$, such that, e.g.,  $c^+(x)dx$ gives the number of filaments
with their plus-end in the positive $x$-direction and their center located in
the interval $[x,x+dx]$.  We assume, that motors are small compared to the
filament length and will be treated as point-like in our description.  The
number density of motors is denoted by $m$.

The filament and motor densities satisfy the conservation laws
\begin{eqnarray}
   \label{eq:cpl} \partial_t c^+ & = & D\partial_x^2 c^+ - \partial_x J^+ \\
   \label{eq:cmi} \partial_t c^- & = & D\partial_x^2 c^- - \partial_x J^- \\
   \label{eq:m} \partial_t m & = & D_\m\partial_x^2 m - \partial_x J
\end{eqnarray}
Here, the currents $J^\pm$ and $J$ are generated by the active interaction
between motors and filaments. The densities $c^+$ and $c^-$ are conserved
separately since we do not allow filaments to change their
orientation. Fluctuations in the system give rise to diffusive terms with
diffusion coefficients $D$ and $D_\m$ of filaments and motors. While the
diffusion of motors  could be expected to result  from thermal fluctuations,
the diffusion of filaments is generated effectively by fluctuations of the
forces induced by motor-filament interactions. For long filaments, the
contribution of thermal fluctuations to the diffusion coefficient  $D$ is
negligible. We return to this point in Sect. \ref{sec:diskussion}.

\subsection{Momentum balance}

In the absence of external forces, the total momentum is conserved in the
filament bundle.  Forces acting within the bundle lead to an exchange of
momentum with the environment or between filaments. Since filaments are
treated as rigid, extended objects with momentum distributed along the full
length of the filament, we introduce the momentum densities
$\pi^\pm(x,y)$. These densities represent the momentum at position $y$ carried
by all plus- or minus-filaments, respectively, with their centers located at
position $x$. The momentum balance can then be expressed as
\begin{widetext}
\begin{eqnarray}
\label{eq:imppl}
\partial_t\pi^+(x,y) + \partial_y\sigma^+(x,y) -f_\text{int}^+(x,y) &  = &
f_{\rm fl}^+(x,y)+f_\m^+(x,y)+f_\text{ext}^+(x,y) \\
\label{eq:impmi}
\partial_t\pi^-(x,y) + \partial_y\sigma^-(x,y) -f_\text{int}^-(x,y) &  = &
f_{\rm fl}^-(x,y)+f_\m^-(x,y)+f_\text{ext}^-(x,y)
\end{eqnarray}
\end{widetext}
Here, momentum flux along filaments centered at $x$ is given by the tensions
$\sigma^\pm(x,y)$. Momentum exchange between filaments is nonlocal and
described by the internal force densities $f_\text{int}^\pm(x,y)$, which
include all active filament interactions via motors. The force densities
$f_\text{fl}^\pm(x,y)$, $f_\m^\pm(x,y)$,  and $f_\text{ext}^\pm(x,y)$ are
source and sink terms, describing momentum exchange with the environment. They
result from friction with the fluid ($f_\text{fl}$), from motors moving along
a single filament ($f_\m$), and from external forces ($f_\text{ext}$). Here,
$x$ refers to filaments with  center at position $x$, while $y$ denotes a
position in space, where a force is acting and momentum is exchanged. Momentum
conservation in the absence of external forces requires, that
\begin{equation}
\label{eq:innereKraefte}
\int dx \left[f^+_\text{int}(x,y) + f^-_\text{int}(x,y)\right] = 0\quad.
\end{equation}
This implies, that any force generated by active cross-links on a filament is
balanced by an opposite force acting on other filaments, i.e., internal forces
at a point $y$ are balanced when integrated  over all filaments.  Therefore,
total momentum $\Pi=\int dx\;dy (\pi^++\pi^-)$ changes according to (ignoring
boundary terms):
\begin{widetext}
\begin{equation}
\label{eq:fb} \frac{d}{dt} \Pi = \int dx\; dy
\left[f_\text{ext}^+(x,y)+f_\text{ext}^-(x,y)+
f_\text{fl}^+(x,y)+f_\text{fl}^-(x,y) +
f_\text{m}^+(x,y)+f_\text{m}^-(x,y)\right]
\end{equation}
\end{widetext}
Inertial terms are negligible in a slowly moving bundle, such that we can set
$\partial_t  \pi^\pm =0$. Equations~(\ref{eq:imppl}) and (\ref{eq:impmi}) then
express a balance of forces.

In the most simple case where friction is local, we can write for the density
of friction forces
\begin{equation}
\label{eq:locfric}
  f^\pm_{\rm fl}(x,y) = \eta J^\pm(x) R(x-y)
\end{equation}
Here, $\eta$ is a friction coefficient per unit length and $R(x)$ is a
function characterizing the distribution of energy dissipation along moving
filaments. If all filaments are of the same length $\ell$, a simple choice is
$R(x)=1$ for $|x|<\ell/2$ and $R(x)=0$ otherwise.  However, the function
$R(x)$ can also account for situations with a distribution of filament
lengths. Then, $R(x)$ is related to the probability, that a given filament is
longer  than $2|x|$.

The forces exerted by motors moving along a single filament are linear in the
filament and motor densities,
\begin{equation}
\label{eq:motfo}
f_\m^\pm(x,y) = \mp \eta_\m \Gamma m(y) c^\pm(x)R(x-y)\quad,
\end{equation}
where $\eta_\m$ is the friction coefficient corresponding to single motors and
the coefficient $\Gamma$ characterizes the binding to and motion on  filaments
of individual motors

\subsection{Currents of filaments and motors}

While the internal forces are balanced at a point $y$ when integrated over all
filaments, the total force $\int dy f_{\rm int }^\pm(x,y)$ acting  on
filaments centered at a given position $x$ does not vanish in general.
Integration of  Eqs.~(\ref{eq:imppl}) and (\ref{eq:impmi}) with respect to $y$
reveals that this force is balanced by friction forces:
\begin{equation}
\label{eq:stromReibung}
\eta\ell J^\pm(x) = -\int dy \left[f^\pm_\text{int}(x,y) + f^\pm_\m(x,y)\right]
\end{equation}
where $\ell = \int dx R(x)$ is the average filament length and where we have
assumed $f_\text{ext}=0$. Since friction of motors is small as compared to
filament friction, $\eta_\m\ll\eta\ell$, the contribution of $f^\pm_\m$ can be
neglected in most practical cases, in particular, for low motor densities. In
the following, we therefore set $\eta_\m=0$.

\begin{figure}[b]
  \includegraphics[width=0.75\linewidth]{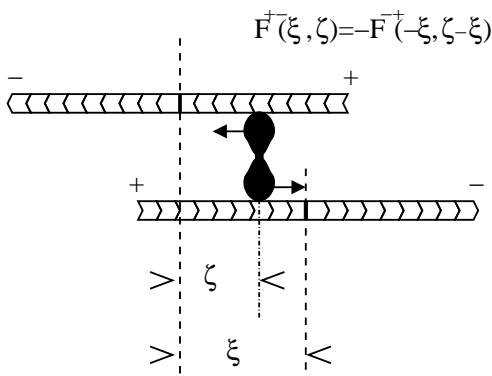} \caption{Schematic
  representation of the forces  
  exerted by an active cross-link on two filaments of opposite
  orientation. The centers of the filaments are indicated by the dashed lines,
  whereas the dotted-dashed line marks the position of the active
  cross-link. The arrows indicate the direction of the forces applied by
  motors on the filaments.}  \label{fig:schema}
\end{figure}

In order to write explicit expressions for the currents, we need a model for
the internal forces in the bundle. We consider the case when clusters of three
or more cross-linked filaments form rarely enough, such that their
contribution to the internal forces can be neglected. This holds in the case
of a low motor density or for motors with a low duty ratio, which is  the
fraction of time a motor spends attached to a filament~\cite{howard97}. If
interactions between filament pairs dominate, we can split the internal forces
into those between filaments of the same and those of opposite orientation. We
write
\begin{equation}
f^+_\text{int} = f^{++}_\text{int} + f^{+-}_\text{int}
\end{equation}
and analogously for $f^-_\text{int}$. A motor may link two filaments and thus
exert forces of opposite sign on each of them, whenever they overlap. Assuming
that the probability for two  filaments  to interact increases  quadratically
with filament density, we write
\begin{widetext}
\begin{equation}
\label{eq:fplpl}
f^{\pm\pm}_\text{int}(x,y) = \int dz\; c^\pm(x)c^\pm(z)\;R(y-x)R(y-z)\;
m(y)F^{\pm\pm}(z-x,y-x)
\end{equation}
\end{widetext}
and corresponding expressions for $f^{\pm\mp}_\text{int}$.  Here,
$F^{++}(\xi,\zeta)$ is the average force acting on plus-filaments at  a
distance  $\zeta$ from the center exerted by motors, that interact with other
minus-filaments located at a distance $\xi$ from the first, see
Fig.~\ref{fig:schema}.  The essential feature of  motor-filament interactions
is that the direction of the force applied by a motor on a filament is
uniquely determined by the orientation of the filament~\cite{ikhhfy96}.  The
product  $R(y-x)R(y-z)$ gives the probability, that a filament at $x$ has an
overlap at $y$ with a filament at $z$. Here, the position of a filament is
given by the position of the filament's center. Analogous expressions hold for
the internal forces between minus-filaments as well as between filaments of
opposite orientation.

The forces $F^{\pm\pm}$ and $F^{\pm\mp}$ obey the following symmetry
relations. Momentum balance demands that under an  exchange of filaments, the
force changes sign:
\begin{eqnarray}
\label{eq:imperh}
F^{\pm\pm}(\xi,\zeta) & = &-F^{\pm\pm}(-\xi,\zeta-\xi)\\ F^{\pm\mp}(\xi,\zeta)
& = &-F^{\mp\pm}(-\xi,\zeta-\xi)\quad
\end{eqnarray}
see Fig.~\ref{fig:schema}. Using relation~(\ref{eq:fplpl}), the internal
forces satisfying the above equations verify Eq.~(\ref{eq:innereKraefte}),
which assures momentum conservation. 
Space inversion symmetry requires
\begin{eqnarray}
F^{++}(\xi,\zeta) & = & - F^{--}(-\xi,-\zeta)\\
\label{eq:spiegel}
F^{+-}(\xi,\zeta) & = & - F^{-+}(-\xi,-\zeta)
\end{eqnarray}

Momentum conservation also determines the non-diffusive motor current  $J$ with
\begin{equation}
-\eta_m J(y)=\int dx [ f_\m^+(x,y)+f^-_\m(x,y) ] \label{eq:J}
\end{equation}
Here the forces of individually bound motors are defined in
Eq.~(\ref{eq:motfo}).

~\\  The continuity Eqs.~(\ref{eq:cpl})-(\ref{eq:m})  for the densities
together with the defining Eqs.~(\ref{eq:stromReibung}) and (\ref{eq:fplpl})
of filament currents and Eq.~(\ref{eq:J}) for the  motor currents provide the
full dynamic equations of active filament bundles. The functions $F^{\pm\pm}$
depend on the details of the motor-filament interactions and could be modified
by further proteins bound to the filaments. However, the large scale behaviors
of the system do not depend on the detailed form of these functions. In the
following sections, we  will therefore make simple choices which obey the
symmetry relations discussed above.

\section{The minimal model}
\label{sec:model1}

The minimal model has been introduced in Ref.~\cite{kj00} as a simple model
for filament dynamics. It can be obtained from the general equations derived
in the previous section by choosing $R(x)=1$ for $\vert x\vert <\ell/2$ and
$R(x)=0$ otherwise.  The forces $F^{\pm\pm}$  are chosen to behave as
\begin{eqnarray}
\label{eq:motfopa}
F^{\pm\pm}(\xi,\zeta) & \sim & \text{sgn}(\xi)  \nonumber\\
\label{eq:motfoap}
F^{\pm\mp}(\xi,\zeta) & \sim &\mp 1 \quad,
\end{eqnarray}
where $\text{sgn}(\xi)=1$ for $\xi>0$ and $-1$ otherwise, which represents the
simplest choice compatible with the symmetry requirements.  Furthermore, we
assume in the minimal model that the motor distribution is homogeneous and its
dynamics can be neglected.

The resulting dynamical  equations are most conveniently expressed in
dimensionless form.  We define $\tilde{x}=x/\ell$ and measure lengths in units
of the filament length $\ell$, and a dimensionless time variable
$\tilde{t}=tD/\ell^2$. Furthermore, we introduce dimensionless densities
$\tilde{c}=c\ell$.  Suppressing the tildes, the dynamic equations of the
minimal model can be written as
\begin{widetext}
\begin{eqnarray}
\label{eq:minmod}
  \partial_t c^+(x) & = \partial^2_x c^+(x) & -\alpha\partial_x\int_0^1 d\xi\;
[c^+(x+\xi)-c^+(x-\xi)]c^+(x)  \nonumber \\ \label{eq:cplind} & & +\beta
\partial_x\int_{-1}^1 d\xi\; c^-(x+\xi) c^+(x) \\ \partial_t c^-(x) & =
\partial^2_x c^-(x) & -\alpha\partial_x\int_0^1 d\xi\;
[c^-(x+\xi)-c^-(x-\xi)]c^-(x)  \nonumber \\ \label{eq:cmiind} & & -\beta
\partial_x\int_{-1}^1 d\xi\; c^+(x+\xi) c^-(x) \quad.
\end{eqnarray}
\end{widetext}
where $\alpha$ and $\beta$ are dimensionless coupling constants characterizing
the strength of the motor forces defined in Eqs.~(\ref{eq:motfoap}).  It
follows from the dynamical equations, that the homogeneous state
$c^\pm(x)=c_0^\pm=const.$ is a stationary solution for all values of the
parameters.

\subsection{Oriented bundles - Contraction} 

If all filaments are of the same orientation, one is left with a single
equation
\begin{equation}
\label{eq:partialC}
\partial_t c(x)=\partial_x^2 c(x)-\alpha\partial_x\int_0^1 d\xi\left[
c(x+\xi)-c(x-\xi)\right] c(x)\quad.
\end{equation}
Here, $c$ represents either $c^+$ or $c^-$, depending on the orientation of
the filaments. This nonlinear integro-differential equation is the most simple
description of the active dynamics of a filament bundle. Many of the basic
physical principles underlying self-organization of filament bundles can
already be discussed using this equation.

\subsubsection{Linear stability}

We consider a system of length $L$ with periodic boundary conditions and study
the stability of the homogeneous state with respect to small
perturbations. Periodic boundary conditions imply that the bundle forms a
ring.  Such rings appear, e.g., in eucaryotic cells, in the late stages of
cell division.  We represent the filament density by a Fourier expansion
\begin{equation}
c(x) = \sum_k c_k{\rm e}^{ikx} \quad , \label{eq:fourier}
\end{equation}
with $k=2\pi n/L$, $n=0,\pm1,\ldots$ and where $c_{-k}=c_k^*$. Up to first
order in the Fourier components $c_k$, the dynamics (\ref{eq:partialC}) reads
\begin{eqnarray}
\label{eq:lambdap}
\frac{d}{dt}c_k & = & -\left(k^2 -2\alpha c_0\left(1-\cos k\right)\right)c_k
\\ & \equiv & \lambda(k)c_k\quad,
\end{eqnarray}
for all $k$. This relation implies that for $\alpha c_0\le k^2/2(1-\cos k)$
the mode $c_k$ decays in time, because then $\lambda(k)\le 0$. It follows that
the most unstable mode is the one corresponding to the smallest non-zero wave
number $k=2\pi n/L$ with $n=1$.  This can be demonstrated using
$(\frac{2\pi}{L})^2/2(1-\cos\frac{2\pi}{L}) \le (\frac{2\pi
n}{L})^2/2(1-\cos\frac{2\pi n}{L})$ for all $n>1$ which can be verified by
induction using the equivalent condition $n^2-n^2\cos(2\pi/L)-1+\cos(2\pi
n/L)\ge 0$.  Therefore, the homogeneous state is linearly stable as long as
$\alpha\le\alpha_c$, where the critical value $\alpha_c$ is determined by
$\lambda(2\pi/L)=0$. Explicitly,
\begin{equation}
  \label{eq:alphaCbetanull} \alpha_c=
  \frac{2\pi^2}{c_0L^2(1-\cos(2\pi/L))}\quad.
\end{equation}

The critical value $\alpha_c$ is positive and decreases with increasing $c_0$
and $L$ (for $L\ge1$). Note, that for bundle sizes $L\ge1$, we have
$0<\alpha_c<\infty$.

\subsubsection{Contracted states}

If the homogeneous state is unstable, the system evolves to an inhomogeneous
steady state. We can calculate this state by numerically solving the dynamic
equations or, in the vicinity of the bifurcation, using a systematic expansion
in Fourier modes.  To third order in $c_1$ the equation for the steady state
$\partial_t c=0$ reads
\begin{equation}
\label{eq:dritteOrdnung}
F(\alpha)c_1 -G(\alpha)|c_1|^2 c_1 = 0 \quad ,
\end{equation}
with $F(\alpha)=\lambda(2\pi/L)$ and $G(\alpha)$ given by
Eqs.~(\ref{eq:Fbetanull}) and (\ref{eq:Gbetanull}), see
Appendix~\ref{app:stoerung}. Note, that $F(\alpha_c)=0$.  Expanding $F$ and
$G$ at $\alpha=\alpha_c$,  we find expressions for the Fourier amplitudes
$c_1$ and $c_2$, given by Eqs.~(\ref{eq:C1}) and (\ref{eq:C2}).  This solution
represents a localized distribution of filaments, i.e., a contracted bundle.

\begin{figure}[t]
  \includegraphics[width=\linewidth]{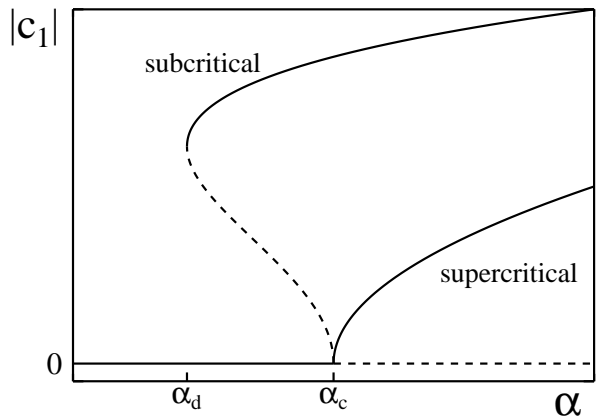} \caption{Schematic
  representation of the 
  amplitude $|c_1|$ of the first spatial Fourier-component of stationary
  states as a function of the coupling strength between equally oriented
  filaments $\alpha$. Presented are the cases of a super- and
  of a subcritical bifurcation. Solid lines represent stable, dashed lines
  unstable solutions. In both cases, the homogeneous state is stable for
  $\alpha<\alpha_c$ and unstable otherwise. In the supercritical case, i.e.,
  if $F/G>0$ for $\alpha>\alpha_c$, the bifurcating solution exists for
  $\alpha>\alpha_c$ and is stable, while in the other case it exists for
  $\alpha<\alpha_c$ and is unstable. In the latter case, one usually finds
  inhomogeneous solutions coexisting with the homogeneous state in
  an interval $[\alpha_d,\alpha_c]$.}  \label{fig:schemaBif}
\end{figure}

It follows from Eq.~(\ref{eq:dritteOrdnung}), that this contracted steady
state exists if $F(\alpha)/G(\alpha)>0$.  Depending on whether the ratio $F/G$
is positive for $\alpha>\alpha_c$ or for $\alpha<\alpha_c$, the bifurcation is
supercritical and subcritical, respectively, see Fig.~\ref{fig:schemaBif}.
From Eq.~(\ref{eq:C1}) one deduces that  the bifurcation is supercritical for
system sizes falling within particular intervals for which
$4/(4n-1)<L<4/(4n-3)$,  $n=1,2,\ldots$.

\begin{figure}[H]
  \includegraphics[width=.99\linewidth]{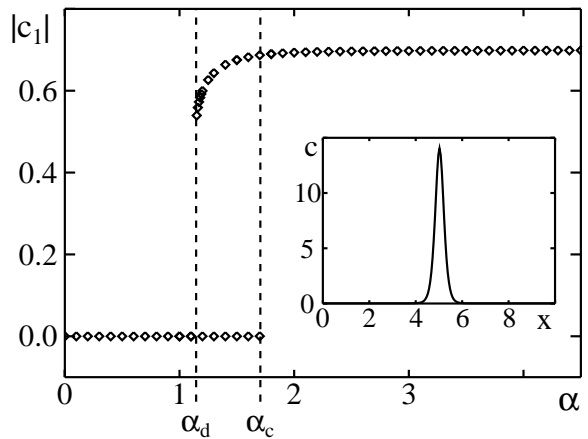} \caption{The amplitude of
  the first 
  Fourier-component of stable stationary solutions of the minimal model for an
  oriented bundle as a function of the interaction strength $\alpha$. The 
  average filament concentration is $c_0=0.7$ and system size $L=10$. The inset
  presents the non-homogeneous stationary solution for $\alpha=1.5\,$. The
  scenario shown corresponds to a subcritical bifurcation, see
  \protect{Fig.~\ref{fig:schemaBif}}.}    \label{fig:attrbnull}
\end{figure}

Figure~\ref{fig:attrbnull} presents numerical solutions of  the dynamical
equations using an Euler-algorithm with spatial discretization
$\Delta=0.1$. Displayed is the modulus of the first Fourier-component $|c_1|$
of the attractors as a function of $\alpha$ for a system of length $L=10$. A
region of coexistence extends from $\alpha=\alpha_d$ up to $\alpha_c$, where
the homogeneous state becomes unstable.

For large $\alpha$ we find numerically, that transient states consisting of
several contracted packets can occur, which however decay for long times to a
steady state with one maximum.  Their lifetime increases with increasing
$\alpha$.

\subsubsection{Contraction dynamics}

The contraction of the bundle is most conveniently discussed in an infinite
system using the variance of the filament distribution
\begin{equation}
  \label{eq:varianz} \sigma^2 = \frac{1}{N}\int_{-\infty}^\infty dx\;
  x^2c(x)\quad,
\end{equation}
where $N=\int_{-\infty}^\infty dx\;c(x)$ is the total number of filaments, as
a measure of bundle contraction.  Since the center of mass of the distribution
is immobile due to momentum conservation,  we have chosen  without loss of
generality $\left\langle x\right\rangle=\int_{-\infty}^\infty dx\;xc(x)=0$.
The variance changes in time as
\begin{widetext}
\begin{eqnarray}
  \label{eq:varianzpunkt} \frac{d}{dt}\sigma^2 & = &
  \frac{1}{N}\int_{-\infty}^\infty\! dx\; x^2 \partial_t c(x,t)\nonumber\\  &
  = & \frac{2}{N}\int_{-\infty}^\infty\! dx\; c(x,t) + \frac{2}{N}\int_0^1\!
  d\xi \left[\int_{-\infty}^\infty\! dx\;x
  c(x+\xi)c(x)-\int_{-\infty}^\infty\!  dx\;x  c(x-\xi)c(x)\right] \nonumber
  \\  & = & 2 - \frac{2}{N}\alpha\int_0^1 d\xi\;\xi \int_{-\infty}^\infty\!
  dx\; c(x+\xi)c(x)\quad.
\end{eqnarray}
\end{widetext}
The final expression reveals two opposing effects.  The positive constant
describes the spreading of the bundle due to fluctuations while the second
term takes into account the effect of the active interactions.  The
interaction between parallel filaments tends to contract the bundle.  Note,
that distributions for which $d\sigma^2/dt=0$ do not necessarily correspond to
stationary solutions of the dynamics (\ref{eq:partialC}).

\subsection{Bundles of mixed orientation - Solitary waves}
\label{sec:betaUnglNull}

\subsubsection{Linear stability}

The linearization of  Eqs.~(\ref{eq:cplind}) and (\ref{eq:cmiind}) around the
homogeneous state $c^\pm(x)=c^\pm_0=const$ reads in Fourier-representation
\begin{equation}
\label{eq:fmmmo}
\frac{d}{dt}\left( \begin{array}{c} c^+_k\\ c^-_k \end{array} \right) = \left(
                    \begin{array}{cc} \Lambda^{++}&\Lambda^{+-}\\
                    \Lambda^{-+}&\Lambda^{--} \end{array}\right )
                    \left(\begin{array}{c} c^+_k\\ c^-_k \end{array}
                    \right)\quad,
\end{equation}
where the elements of the matrix $\Lambda(k)$ are given by
\begin{eqnarray}
\Lambda^{\pm\pm}(k) & =& -k^2-2\alpha(\cos(k)-1)c^\pm_0\pm 2i\beta k c^\mp_0
\nonumber \\ \Lambda^{\pm\mp}(k) & =& \pm 2i\beta\sin(k)c^\pm_0 \quad .
\end{eqnarray}
For a system of length $L$ with periodic boundary conditions, the wave numbers
are $k=2\pi n/L$ with $n=0,1,\ldots$.  In presence of the coupling between the
filaments of opposite orientation, the matrix $\Lambda$ is not diagonal.  The
stability of the modes is determined by the larger of the real parts
$\lambda(k)$ of the complex eigenvalues of this matrix which are given in
App.~\ref{app:eigenwerte}.

We find again that the mode with the smallest wave number $k=2\pi/L$ is most
unstable and there exists a critical value $\alpha_c$ where the homogeneous
state becomes linearly unstable.  Furthermore, $\alpha_c\ge0$, independently
of the values of the other parameters, see App.~\ref{app:eigenwerte}.   The
critical value
\begin{equation}
\alpha_c \equiv \frac{1}{c}g\left(\beta, \frac{\delta c}{c}, L\right),
\end{equation}
is a function of the remaining parameters, where $c=c^+_0+c^-_0$ and $\delta
c=c^+_0-c^-_0$. Here, $g$ is a dimensionless scaling function. In some
limiting cases, explicit expressions for $\alpha_{c}$ can be obtained.  For
example, for $\delta c=0$ one finds $g=4\pi^2/L^2[1-\cos(2\pi/L)]$ and in the
limit $L\rightarrow\infty$ $g=1$ if $\beta\neq0$, whereas $g=2/(1+|\delta
c|/c)$ for $\beta=0$, see Eq.~(\ref{eq:alphaCbetanull}).  These expressions
reflect some interesting properties of $\alpha_c$, e.g., it decreases
monotonically with $|\delta c|/c$, $L$, and $|\beta|$ \cite{kj00}.

\subsubsection{\label{sec:solitaires} Solitary waves}

For $\beta\neq0$, the eigenvalues of $\Lambda(k)$ are complex and the
homogeneous state loses stability through a Hopf-bifurcation, leading to
solutions that oscillate in time. We find that at the bifurcation a solitary
wave of the form  $c^\pm(x,t)=u^\pm(x-vt)$ occurs.  From momentum
conservation, it follows that the total filament current
\begin{equation}
I=\int dx\; (J^{+}+J^{-})
\end{equation}
vanishes. This implies that the propagating filament pattern is not
accompanied by a net filament transport. However, as soon as filament
adhesion to a substrate is introduced, the total filament current associated
with a solitary wave no longer vanishes and self-organized filament transport
occurs~\cite{kcj01},  see App.~\ref{sec:adhaesion}.

For weak interactions between filaments of opposite orientation,
$|\beta|\ll1$, solitary waves can be understood intuitively. They emerge from
the interaction of a contracted distribution of filaments of one orientation
with a homogeneous distribution of filaments of opposite orientation.  This
picture suggests a systematic procedure for determining solitary
waves. Writing $c^\pm(x,t)=u^\pm(x-vt)$, we can expand $u^\pm(x)$ in  powers
of $\beta$. For $\beta=0$, we start from steady states as discussed above,
which we denote as $c^\pm(x,t)=u_{0}^\pm (x)$. Solitary waves are obtained by
assuming that, e.g., $u_0^+$ is a contracted steady state, while $u_{0}^-$ is
homogeneous.  We can now write
\begin{eqnarray}
\label{eq:ansatzC}
u^\pm(x) &=& u^\pm_0(x) + u^\pm_1(x)\beta+u^\pm_{2}\beta^2 + \ldots \\
\label{eq:ansatzV}
v &=& v_{1}\beta + v_{3}\beta^3 + \ldots
\end{eqnarray}
The even terms in the expansion for $v$ vanish by symmetry. To  lowest order,
we obtain
\begin{eqnarray}
\label{eq:erstord1}
v_1 & = & 2c_0^- \\
\label{eq:erstord2}
u_1^+ & = & 0\\
\label{eq:erstord3}
u_{1,k}^- & = &  -\frac{2i\sin k\; c^-_0 u^+_{0,k}} {k^2-2\alpha(1-\cos k)
c^-_0}\quad {\rm for}\, k\neq 0. \label{eq:Ukminus}
\end{eqnarray}
This result agrees with numerical solutions of Eqs.~(\ref{eq:cplind}) and
(\ref{eq:cmiind}), see Ref.~\cite{kcj01}.  For any $\beta$, solitary wave
solutions can be obtained near the instability via a systematic expansion in
Fourier modes, see App.~\ref{app:betaUnglNull}.

\subsubsection{Bifurcation diagrams}

\begin{figure}[t]
  \includegraphics[width=\linewidth]{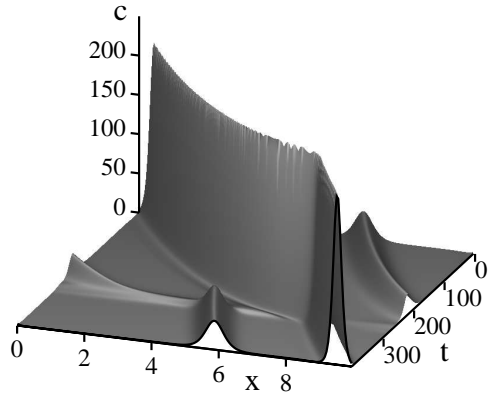} \caption{Total filament
  concentration $c=c^++c^-$ 
  as a function of position $x$ and time $t$ for An oscillating wave solution
  of the minimal model for coupling parameters $\alpha=2.5$ and $\beta=1.$,
  average filament densities $c^+_0=0.7$ and $c^-_0=0.3$,
  and system size $L=10$.}  \label{fig:oszwelle}
\end{figure}

The arguments of the previous subsection suggest a general scenario of
bifurcations as the interaction strength $\alpha$ is increased. This scenario
emerges from the situation with $\beta=0$, where the two subpopulations evolve
independently, if $\beta$ becomes nonzero. For small $\alpha$ the homogeneous
state is stable. At the critical value $\alpha_c$, corresponding to the
instability of one filament population at $\beta=0$, a solitary wave occurs.
Furthermore, a second bifurcation where the solitary wave loses its stability
occurs at $\alpha_c'$. This second bifurcation is related to the point where
the second filament population becomes unstable for $\beta=0$. This
bifurcation leads to an oscillating wave solution which consists of two
oscillating distributions, which periodically  penetrate each other. An
example of such an oscillating wave is shown in Fig.~\ref{fig:oszwelle}. It is
characterized by propagating filament profiles which oscillate.

\begin{figure}[t]
  \hspace*{-2.7cm}\includegraphics[width=1.75\linewidth]{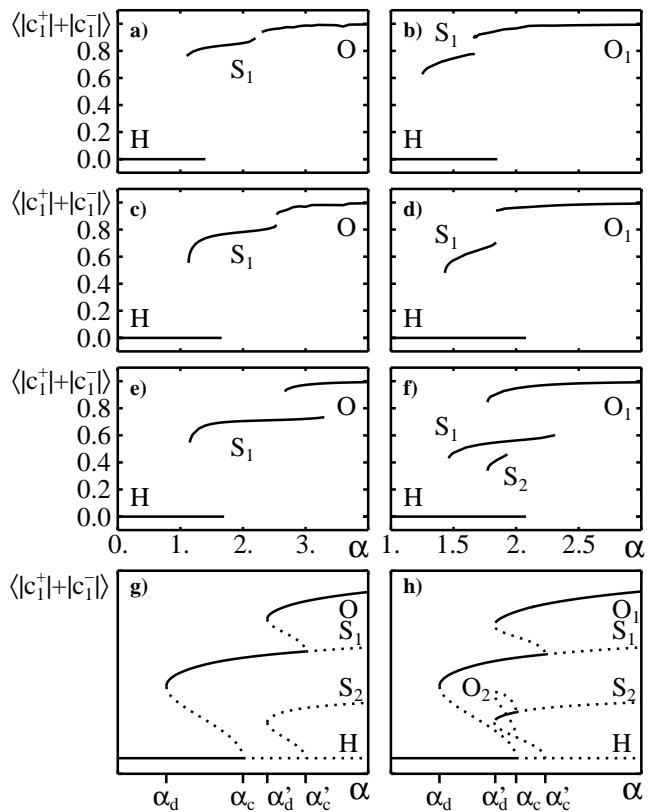}
  \caption{Asymptotic solutions of the minimal 
  model represented by the time-averaged total amplitude $|c^+_1|+|c^-_1|$ of
  the first spatial Fourier-components as a function of the interaction
  strength$\alpha$ for different values of the interaction strength
  $\beta$. The homogenous state is indicated by H, solitary waves by S, and
  oscillatory waves by O.  (a)-(f) have been obtained by numerical integration
  of the minimal model, where $\beta=0.1$ for (a) and (b), $\beta=0.01$ for
  (c) and
  (d), and $\beta=0.001 for (e)and (f)$. (a), (c), (e) are for
  $c_0^+=0.7$ and $c_0^-=0.3$, while (b), (d), (f) are for $c_0^+=0.55$ and
  $c_0^-=0.45\;$. In all cases $L=10$. (g) and (h) give a schematic
  representation valid for $|\beta|\ll1$ derived from the bifurcation diagram
  for $\beta=0$ as explained in the text. Dashed lines indicate unstable
  solutions.}  \label{fig:attraktoren}
\end{figure}

These arguments allow us to derive, the full bifurcation scenario for small
values of $\beta$.  For $\alpha_{d}<\alpha<\alpha_{c}$ homogeneous
distributions and solitary waves coexist.  Similarly, we find a coexistence of
oscillating and solitary waves for $\alpha_d'<\alpha<\alpha_c'$.  Depending on
the ratio of plus- and minus-filament numbers, $c_0^+/c_0^-$, we find
different bifurcation scenarios, see Fig.~\ref{fig:attraktoren}. For large
ratios, $\alpha_c<\alpha_d'$ and the coexistence regions are separated. In the
second case of similar filament  numbers  shown in Fig.~\ref{fig:attraktoren},
$\alpha_d'<\alpha_c$ and multiple coexistence occurs, in particular a
coexistence of two different solitary waves indicated as $S_1$ and
$S_2$. These waves move into opposite directions. For the special case
$c_0^+=c_0^-$, the system is  symmetric under $x\rightarrow -x$ and the
coexisting waves $S_1$ and $S_2$ occur via spontaneous symmetry breaking.

For three different values of $\beta$, numerically obtained bifurcation
diagrams are displayed in Fig.~\ref{fig:attraktoren}. Shown is $\left\langle
|c_1^+| +|c_1^-|\right\rangle$, i.e., the time-averaged sum of the amplitudes
of the first spatial Fourier components of both distributions as a function of
$\alpha$.   For $\beta=0.001$ the diagram follows closely the curves shown in
the bottom panel corresponding to the limit of small $\beta$.  As $\beta$ is
increased, the coexistence regions shrink and the solitary wave $S_2$
disappears. The bifurcation from solitary waves to oscillating waves becomes
supercritical for large $\beta$.

\section{\label{sec:model2} Dynamic motor distributions}

Motors are actively transported along filaments which in general leads to
dynamic changes in the motor distribution \cite{nsm01,lk01}. In the previous
section, we have assumed that the motor distribution remains homogeneous,
which implies that motors diffuse infinitely fast. We will now discuss the
effects of the dynamics of the motor distribution using the same choices for
the motor forces as given by Eqs.~(\ref{eq:motfoap}) and we assume that
filaments are of the same length $\ell$.

\subsection{Dynamical equations}

The dynamic equations, including the dynamics of the motor density, have been
derived in section II. As in the last section, we use  dimensionless space and
time coordinates, $\tilde{x}=x/\ell$ and $\tilde{t}=tD/\ell^2$ as well as
dimensionless densities $\tilde{c}^{\pm}$. We introduce a dimensionless motor
density $\tilde{m}=m/m_0$, where $m_0=\frac{1}{L}\int_0^Ldx\;m(x)$ and $L$ is
the system size. Furthermore, we define the dimensionless parameters
$\tilde{\Gamma}=\Gamma\ell/D$, and $\tilde{D}_\m=D_\m/D$. Suppressing the
tildes we obtain
\begin{eqnarray}
  \label{eq:model2cpm} \partial_t c^\pm & = & \partial_x^2c^\pm -
  \partial_xJ^{\pm\pm} - \partial_xJ^{\pm\mp} \\ \label{eq:model2cm}
  \partial_t m & = & D_\m\partial_x^2m - \partial_xJ
\end{eqnarray}
with
\begin{eqnarray}
  \label{eq:model2strompa}  J^{\pm\pm}(x) & = & A\!\int_{-1}^1\!d\xi\;
  \text{sgn}(\xi)\;M(x,\xi) c^\pm(x+\xi) c^\pm(x)\\  \label{eq:model2stromap}
  J^{\pm\mp}(x) & = & \mp B\!\int_{-1}^1\!d\xi\;
  M(x,\xi)c^\mp(x+\xi)c^\pm(x)\\ \label{eq:model2strommo}  J(x) & = &
  \Gamma\!\int_{-\frac{1}{2}}^{\frac{1}{2}}\!d\xi
  \left[c^+(x+\xi)-c^-(x+\xi)\right] m(x) \quad.
\end{eqnarray}
Here $M$ is the number of motors present in the overlap

\noindent region of two
filaments with  $M(x,\xi)=\intpl m(\zeta)$ for $\xi>0$ and $M(x,\xi)=\intmi
m(\zeta)$ for $\xi<0$. Furthermore, the dimensionless coupling constants $A$
and $B$ are related to the parameters $\alpha$ and $\beta$ of the minimal
model and describe effective interactions of a motor with a filament pair.

\subsection{Oriented bundles}

\begin{figure}[H]
  \vspace*{-.5cm}\hspace*{2cm}\includegraphics[width=1.3\linewidth]{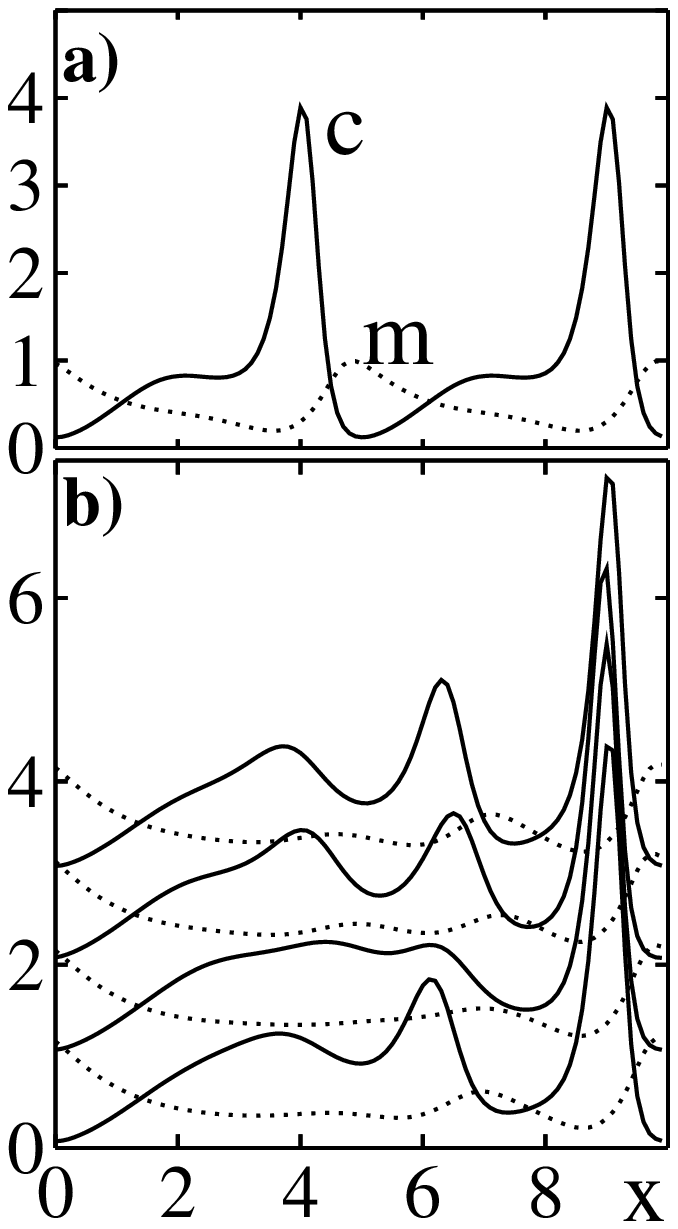}
  \label{fig:attraktorKomp}
\end{figure}

We consider again a system of length $L$ with periodic boundary conditions and
all filaments of the same orientation characterized by the density $c(x)$.
The dynamic equations linearized at the homogeneous state with $c(x)=c_0$ and
$m(x)=m_0=1$ read in Fourier representation
\begin{widetext}
\begin{equation}
\label{eq:linmot1}
\frac{d}{dt} \left(
\begin{array}{c}
c_k\\ m_k
\end{array} \right )
= \left( \begin{array}{cc} -k^2+2A(1-\frac{\sin k}{k})c_0 &
-2A(\cos\frac{k}{2}-\frac{2}{k}\sin\frac{k}{2}){c_0}^2  \\
-2i\Gamma\sin\frac{k}{2} & -D_\m k^2 - i\Gamma k c_0 \end{array}\right)  \left(
\begin{array}{c}
c_k\\ m_k
\end{array} \right )
\quad,
\end{equation}
\end{widetext}
where $k=2\pi n/L$, with $n$ being integer.  For $\Gamma/D_\m\rightarrow0$, we
recover the minimal model discussed in the previous section.  In contrast to
the minimal model, the eigenvalues of~(\ref{eq:linmot1}) are complex if
$\Gamma\neq0$. The homogeneous state thus loses stability through a
Hopf-bifurcation, where the most unstable mode occurs at a characteristic
wave-number $k_0$.  As a consequence, for larger system sizes, solitary waves
with multiple maxima appear, see Fig.~\ref{fig:attraktorKomp}.  Furthermore,
oscillatory waves which can coexist with solitary waves have been found, in
contrast to oriented bundles in the minimal model.

\subsection{Bundles of mixed orientation}

In the case of mixed bundles, the linearized equations read in Fourier
representation
\begin{equation}
\frac{d}{dt}\left( \begin{array}{c} c^+_k\\ c^-_k \\ m_k \end{array} \right) =
                    \left( \begin{array}{ccc} \Lambda^{++} & \Lambda^{+-} &
                    \Lambda^{+\m}\\ \Lambda^{-+} & \Lambda^{--} &
                    \Lambda^{-\m}\\ \Lambda^{\m+} & \Lambda^{\m-} &
                    \Lambda^{\m\m} \end{array}\right ) \left(\begin{array}{c}
                    c^+_k\\ c^-_k\\ m_k \end{array} \right),
\end{equation}
where the elements of the matrix $\Lambda(k)$ are
\begin{eqnarray}
\Lambda^{\pm\pm}(k) & =& -k^2+2A(1-\frac{\sin k}{k}) c^\pm_0\pm iB k c^\mp_0
\nonumber\\  \Lambda^{\pm\mp}(k) & =& \pm 2iB\frac{1-\cos k}{k}c^\pm_0
\nonumber\\ \Lambda^{\pm\m}(k) & = &
-2A(\cos\frac{k}{2}-\frac{2}{k}\sin\frac{k}{2}) {c_0^\pm}^2 \pm
2iB\sin\frac{k}{2}c_0^+c_0^- \nonumber\\  \Lambda^{\m\pm}(k) & = & \mp2i\Gamma
\sin \frac{k}{2} \nonumber\\  \Lambda^{\m\m}(k) & = & -D_\m k^2 -i\Gamma
k(c_0^+-c_0^-)\nonumber
\end{eqnarray}
\vspace*{-.5cm}
\begin{figure}[H]
 \caption{Examples of asymptotic
  solutions for an 
  oriented bundle, where the dynamics of the motor distribution has been taken
  into account. (a) shows a solitary wave with a spatial period of half the
  system size, (b) an oscillatory wave. The solid lines represent the filament
  distributions, the dashed lines the motor distributions. (b) distributions
  at four different times are shown. They have been shifted in the
  $y$-direction for better visibility and in the $x$-direction such that the
  positions of the maxima coincide. The two solutions in (a) and (b) exist for
  the same parameters, which are $A=1.5$, $\Gamma=1.5$, $D_\m=1.$,
  $c_0=1.$, $m_0=0.5$, and $L=10.$ }   
\end{figure}

\begin{figure}[t]
  \hspace*{1.1cm}\includegraphics[width=1.5\linewidth]{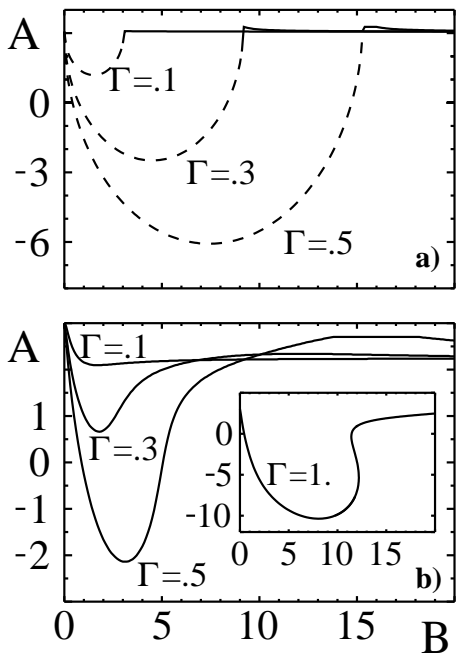} \caption{Regions of stability
  of the homogeneous 
  state for a filament-bundle with dynamic motor distribution. $A$ and $B$
  characterize the coupling of a filament pair via a motor, where the
  homogeneous state is stable below the lines shown. (a) is for the symmetric
  case $c_0^+=c^-_0=1$, (b) for $c_0^+=1$ and $c_0^-=0.5\;.$ In both cases,
  $L=10$ and $D_\text{m}=1$. Solid lines indicate oscillatory instabilities,
  dashed lines static instabilities. The inset presents a case where
  increasing $A$ can re-stabilize the homogeneous state.}
  \label{fig:linstabKomp}
\end{figure}
Fig.~\ref{fig:linstabKomp} displays  the line in the $(A,B)$-plane limiting
the region of stability of the homogeneous state for different values of
$\Gamma$. In contrast to the minimal model, it depends non-monotonically on
$B$. This implies that upon increasing $B$ the homogeneous state can be
re-stabilized. For large enough $\Gamma$ we even find reentrant behavior with
respect to $A$, i.e.,  increasing $A$ can re-stabilize the homogeneous state,
see Fig.~\ref{fig:linstabKomp}. Furthermore, the figure shows that the
critical values of $A$ can become negative. This implies that the homogeneous
state can become unstable even if interactions between parallel filaments are
absent i.e. $A=0$. Another interesting observation is that for  $c_0^+=c_0^-$
the homogeneous state becomes unstable with respect to stationary
inhomogeneous states in certain parameter ranges of $B$.  These inhomogeneous
stationary states represent a new type of solution as compared to the minimal
model of bundles of mixed orientation.

\begin{figure}[t]
  \includegraphics[width=\linewidth]{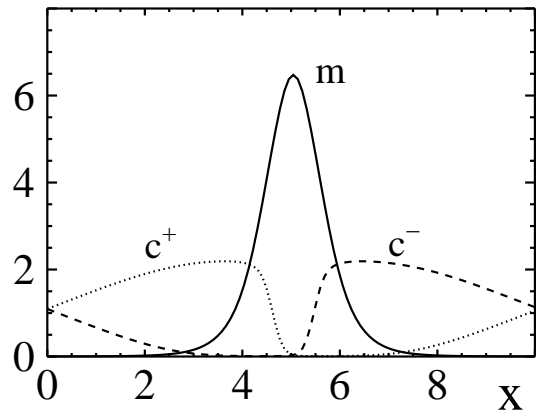} \caption{Inhomogeneous steady state of a
  symmetric system with $c^+_0=c^-_0$ where the dynamics of the motor
  distribution $m$ has been taken into account.  The filament distributions
  are represented by dotted and dashed lines for $c^+$ and $c^-$,
  respectively, the full line represents the motor density $m$. The parameters
  are $c^+_0=c^-_0=1$, $B=\Gamma=D_\m=1$, and $L=10$.}  \label{fig:1da}
\end{figure}

Dynamic solutions to this model can be studied numerically.  We find the three
types of solutions discussed before namely  homogeneous states, solitary
waves, and oscillatory waves.  In addition we obtain  the inhomogeneous
stationary state mentioned above  which is a one-dimensional analog of asters
in higher dimensions and consist of localized distributions of motors and
filaments, see Fig.~\ref{fig:1da}.  In contrast to the minimal model, solitary
waves can exist even in the absence of interactions between filaments of
opposite orientation ($B=0$) and stationary inhomogeneous states can exist for
$B\neq 0$.  Interestingly, the direction of motion of solitary waves can be
reverted by changing $\Gamma$.

~\\ In summary, the important effects of a dynamic motor distribution are (i)
the appearance of a characteristic wave-length independent of the system size,
(ii) the appearance of aster-like solutions, and (iii) that the existence of
inhomogeneous states requires a symmetric situation with $c_0^+=c_0^-$.

\section{\label{sec:spannung}Contractile tension and external forces}

So far, we have been focusing on dynamical properties of active bundles.  We
now discuss mechanical properties of the bundle such as bundle tension and the
role of applied external forces.  The  total tension $\Sigma(y)$ at a point
$y$ within the bundle is obtained by integrating the contributions of all
filaments, i.e.,
\begin{equation}
\Sigma(y) = \int dx \left[\sigma^+(x,y)+\sigma^-(x,y)\right]\quad,
\end{equation}
where $\sigma^\pm(x,y)$ has been introduced in Eqs.~(\ref{eq:imppl}) and
(\ref{eq:impmi}). Taking into account momentum conservation,
Eq.~(\ref{eq:innereKraefte}), one finds in the absence of external forces
\begin{equation}
\label{eq:sigmastrich}
\frac{d}{dy}\Sigma(y) = \int dx\left[f^+_\text{fl}+f^-_\text{fl}+f^+_\m +
f^-_\m\right]\quad.
\end{equation}
This equation allows us to calculate the tension profile by means of
Eq.~(\ref{eq:locfric}) if the currents $J^\pm$ are known.

For simplicity, we neglect again $f_\m$. The total tension in the bundle can
be written as
\begin{equation}
\Sigma(y) = \Sigma_\rightrightarrows(y)+ \Sigma_\rightleftarrows(y)\quad.
\end{equation}
Here, we have introduced the tension due to interactions  between filaments of
the same orientation, $\Sigma_\rightrightarrows$, and those of opposite
orientation, $\Sigma_\rightleftarrows$.  From Eq.~(\ref{eq:sigmastrich}) we
obtain
\begin{widetext}
\begin{eqnarray}
\label{eq:sigmapa}
\Sigma_\rightrightarrows(y) & = &-\eta\int dx
(J^{++}(x)+J^{--}(x))Q(x-y) %\nonumber\\ 
+ \Sigma_\rightrightarrows^{(0)}\\
\label{eq:sigmaap}
\Sigma_\rightleftarrows(y) & = &-\eta\int dx
(J^{+-}(x)+J^{-+}(x))Q(x-y) %\nonumber\\ 
 + \Sigma_\rightleftarrows^{(0)}\quad,
\end{eqnarray}
\end{widetext}
where $\frac{d}{dx}Q(x)=R(x)$ with $Q(0)=0$, and
$\Sigma_\rightrightarrows^{(0)}$ and $\Sigma_\rightleftarrows^{(0)}$ are
constants of integration. These expressions seem to imply that the tension at
position $y$ depends on the global state of the bundle since $Q(x)\neq 0$ for
all $x\neq 0$.   However, tension is a local quantity as can be seen by
introducing  the function $P(x,\xi) = Q(x+\xi)-Q(x)$. For the minimal model
with $R(x)=1$ for $\vert x\vert <\ell/2$ and $R(x)=0$ otherwise we obtain
\begin{widetext}
\begin{eqnarray}
\label{eq:spa}
\Sigma_\rightrightarrows(y) & = & \frac{1}{2}\alpha\eta \int_{-\ell}^\ell d\xi
\int dx\;c^+(x+\xi)c^+(x) P(x-y,\xi)\text{sgn}(\xi) 
\nonumber\\ & &
+\frac{1}{2}\alpha\eta \int_{-\ell}^\ell d\xi \int dx\;c^-(x+\xi)c^-(x)
P(x-y,\xi)\text{sgn}(\xi)
\end{eqnarray}
and
\begin{eqnarray}
\label{eq:sap}
\Sigma_\rightleftarrows(y) & = & \frac{1}{2}\beta\eta \int_{-\ell}^\ell d\xi
\int dx\;c^-(x+\xi)c^+(x) P(x-y,\xi) 
\nonumber\\ & & 
-\frac{1}{2}\beta\eta
\int_{-\ell}^\ell d\xi \int dx\;c^+(x+\xi)c^-(x) P(x-y,\xi)\quad.
\end{eqnarray}
\end{widetext}
In contrast to the non-local function $Q$, $P$ is a local function, with
$P(x,\xi)=0$ for $|x|>2\ell$. In writing the above equations, we have dropped
constant contributions to the tension arising from the boundaries. Indeed, if
we consider a situation with periodic boundary conditions, such boundary terms
must vanish. The integration constants $\Sigma_\rightrightarrows^{(0)}$ and
$\Sigma_\rightleftarrows^{(0)}$ are therefore determined by the condition that
in the final expressions (\ref{eq:spa}) and (\ref{eq:sap}) constant
contributions to the tension are absent.

The tension in the homogeneous state $c^\pm(x) = c_0^\pm$, is given by

\begin{equation}
\Sigma = \frac{1}{2}\eta\ell^3\alpha({c_0^+}^2+{c_0^-}^2) \quad.
\end{equation}
In this case only the interaction between parallel filaments contributes to
the active part of the bundle tension. For an oriented bundle, the tension is
positive, i.e., contractile, whenever $\alpha>0$.

More compact expressions can be obtained by the approximation $P(x,\xi)=\xi$
for $|x|<\ell/2$ and $P(x,\xi)=0$ elsewhere. Then
\begin{widetext}
\begin{equation}
\Sigma_\rightrightarrows(y) \approx
\frac{1}{2}\alpha\eta\int_{y-\ell/2}^{y+\ell/2}dx \int_{-\ell}^\ell d\xi\;
|\xi|(c^+(x+\xi)c^+(x)+c^-(x+\xi)c^-(x))
\end{equation}
and
\begin{equation}
\Sigma_\rightleftarrows(y) \approx
\frac{1}{2}\beta\eta\int_{y-\ell/2}^{y+\ell/2} dx \int_{-\ell}^\ell d\xi\;\xi
(c^+(x+\xi)c^-(x)-c^-(x+\xi)c^+(x))\quad.
\end{equation}
\end{widetext}
For filament distributions that vary weakly over a filament length, this
result corresponds to the expressions given in Ref.~\cite{kj00}, with
$\bar\eta=\eta/2$ and $\tilde\eta=\eta/4$.

Contractile tension in the bundle can give rise to contractile forces exerted
by the  bundle. In order to illustrate this, consider a homogeneous oriented
bundle  with constant filament  density $c(x)=c_0$ inside a box $0\le  x\le L$
of size $L$ while $c(x)=0$ elsewhere. In order to stabilize this state, we
impose boundary conditions which immobilize filaments within the intervals
$[0,\ell]$ and $[L-\ell,L]$ near the ends. Such boundary conditions could be
realized by attaching the filaments near the end to a substrate. This filament
distribution is stationary and for $\alpha<\alpha_c$ stable in the interval
$[\ell, L-\ell]$. At the ends, the force balance Eq.~(\ref{eq:fb}) is
satisfied only if the force density $f_\text{ext}(x) = \alpha\eta\ell
c_0^2(\ell-x)$ is applied for $x\in[0,\ell]$ and correspondingly at the right
end. The total force acting on the ends is thus $F=\int_0^\ell
f_\text{ext}(x)dx=\Sigma$. This result indicates, that the generated force is
independent of the bundle length and increases with the square of the filament
density.

\section{\label{sec:diskussion}Discussion}

In this paper, we have developed a physical description for the dynamics and
mechanics of active filament bundles. In this one-dimensional description, we
describe the dynamics of the filament densities projected on an axis which is
parallel to the filaments. The activity in these bundles results from active
cross-links, that create relative forces between the filaments. Our approach is
based on momentum conservation  within the bundle and momentum  exchange via
external forces. Dynamic equations can be derived most conveniently using
simplifying assumptions such as a low motor density or low duty ratio, local
friction, the absence of passive cross-linkers, and the assumption that
filaments do not change their lengths.  At low motor densities the filament
currents are dominantly generated  by interactions of filament pairs. Using
this approach we  systematically derive the minimal model introduced in
earlier works.  In general, the filament dynamics in the active bundle is
described by nonlinear integro-differential equations. The nonlocal character
of these equations reflect the finite filament length.

The bundle dynamics on scales much larger than the filament length
alternatively can be described phenomenologically in a continuum limit,
leading to a non-linear description in the form of partial differential
equations~\cite{kjup}. Such an approach is not limited by the above mentioned
simplifying assumptions. However, the origin of the different terms which
appear in the equations cannot be systematically related to more microscopic
mechanisms. Interestingly,  the main features found  in phenomenological
descriptions are already captured by the  more specific models described here,
which are derived using simple approximations.  This suggests that the main
features obtained from our models can still hold in situations where our
approximations are no longer valid.

The equations discussed in this paper  describe the average behavior of the
bundle and thus represent a mean-field theory where fluctuations are captured
by effective diffusion terms but do not explicitly appear in the
description. In particular, the coefficient $D$ of filament diffusion is
effectively generated by motor-filament interactions. In the absence of such
interactions, thermal fluctuations would lead to a diffusion coefficient
$D\simeq kT/\eta \ell$, which becomes small for long filaments.  The effective
diffusion coefficient due to active interactions between parallel filaments,
which do not generate a net current in the homogeneous state, can be estimated
as $D\simeq \Delta x^2 \omega$. Here $\Delta x$ is the run-length of a motor
on a filament and $\omega\sim \ell$ denotes the rate of generation of mobile
cross-links, which grows linearly with the filament length. As  a consequence,
$D\sim \ell$ and for long filaments the diffusion is dominated by active
cross-links.  Numerical simulations of computer models, which take into
account fluctuations show, that the phenomena described in the previous
sections persist qualitatively in the presence of fluctuations~\cite{kj00}. A
thorough analysis of the effect of fluctuations will be the subject of a
separate publication~\cite{kjup}.

The minimal model which neglects the dynamics of the motor distribution
already exhibits a complex scenario of behaviors. We discuss the full
bifurcation diagram of this model which involves homogeneous solutions,
contracted steady states, solitary waves and oscillatory waves. These states
are separated by bifurcations and can partly coexist depending on parameter
values.

Taking into account the dynamics of the motor distribution does again generate
the types of solutions present in the minimal model. Furthermore, new
inhomogeneous steady states occur which are the one-dimensional analog of
asters.  In addition, the bifurcation diagrams are modified and are
considerably richer.  Contracted steady states are destabilized  by the motor
dynamics and in many cases become solitary waves. Furthermore, instabilities
of the homogeneous state already occur in the absence of interactions between
filaments of the same orientation, i.e. $A=0$. The bifurcation diagrams
exhibits for certain parameter ranges  reentrant behavior, i.e. the
homogeneous state is re-stabilized by increasing $A$ or $B$.

External forces modify the filament dynamics in interesting ways.
Using the momentum balance in the filament bundle, we studied
active mechanical properties of the bundle. In general, the bundle generates
mechanical tension. Applying external forces to the bundle ends, a stationary
state of a contractile bundle can be attained.

It is interesting to compare our results  to experiments, where purified
filaments interact with motor molecules or small aggregates of such
motors. There is one in vitro experiment involving filament
bundles~\cite{takiguchi91}, where the contraction accompanied by polarity
sorting of disordered bundles of actin filaments in the presence of ATP and
myosin II molecules (more specifically HMM), which presumably spontaneously
form  small aggregates, has been observed. Qualitatively, this corresponds to
the state shown in Fig.~\ref{fig:1da}. However, this experiment has not been
repeated and a systematic in vitro study exploring all the regimes discussed
in this paper is lacking and would be very valuable.  Experiments on filament
bundles could also be performed using microtubules, however, suitable
preparations techniques to generate aligned filaments have to be
developed. Artificially constructed kinesin aggregates would be natural
candidates for mobile cross-links used in such a study.

As mentioned above, our results are expected to be more general and could also
apply to situations where additional components are present. For example it
may be more convenient to prepare filament bundles using the help of passive
cross-linking or bundling proteins like $\alpha$-actinin. For a large
cross-linker density or a long lifetime of such passive cross-links we expect
qualitatively new behaviors. However, for low concentrations or short
lifetimes of passive cross-links they are expected to mainly  modify the
effective friction coefficient and possibly other parameters of our
model. Even in the presence of passive cross-links, we expect the main results
of our work to apply in certain regimes.

A similar modification of model parameters can be expected in general, if
other proteins are present which interact with motors and/or filaments.  For
example, the interaction strength $\alpha$ between equally oriented filaments
could be significantly enhanced by proteins bound to filaments which affect the
speed of motors. In fact, it might seem odd at first glance that filaments of
the same orientation exhibit significant interactions via motors at all.  As a
motor advances on both filaments with the same speed, no relative motion is
generated.  Interactions between filaments of the same orientation are induced
by motors that do not move with the same speed on two cross-linked
filaments. This happens, e.g., when a motor arrives on one filament at the end
towards which it moves. In this case, the motor stops on one filament while
continuing for a while  to move on the second filament. This induces relative
filament sliding via an end-effect.

The interaction strength $\alpha$ 
between equally oriented filaments can be enhanced if the motor speed varies
along the whole filament. 
For example, the speed of motors can be affected by the presence of other
motors on the same filament. Such crowding would typically lead to
a slowing down of motors as they approach the filament end~\cite{ks02},
generating relative motion of filament pairs. 
Furthermore, one could imagine specific proteins bound to the filaments which affect
the speed of motors. If such proteins had a graded distribution linked to
the filament polarity, they would lead to strong interaction terms between filaments of
the same orientation. These examples illustrate, that in more complex 
situations where additional proteins are present, the main results of our work
could still apply, however, with effective parameters. 

This suggest, that the essential properties of active filament bundles found
in our work might be even more general and could also apply to more complex
situations found in vivo.  For example, our results could apply to stress
fibers. These are contractile actin bundles in cells lacking the obvious
periodic organization of muscles~\cite{ajlrrw02} but containing myosins and
other proteins.  As we have demonstrated in Sect.~\ref{sec:spannung}, the
generation  of tension and contraction is possible through the interaction of
filament pairs without the need of a muscle-like sarcomere structure. The
periodic boundary conditions which we use in several examples correspond to
the situation, where the bundle forms a ring and could apply for example to
contractile rings which cleaves a cell during cell division.

Interestingly, the types of dynamic behaviors which we observe also include
qualitatively the symmetry breaking presented by fragments of fish
keratocytes~\cite{es84,vsb99}.   These fragments consist of the lamellipodium,
which is the flattened leading margin of these cells, responsible for their
migration. Notably, they do contain neither the nucleus nor
microtubules. These fragments exist in a symmetric stationary state as well as
in an asymmetric locomoting state, where one can change between these states
through sufficiently strong external perturbations~\cite{vsb99}. Even though
the  active bundles studied in the present work are far from giving a
description of moving keratocyte fragments, our results clearly indicate, that
viewing the cytoskeleton as a dynamical system is a valuable concept  for
understanding such phenomena. Our description of active bundles provides a
firm basis for the development of  more profound theories of active filament
systems, which could help understanding self-organization and dynamic
behaviors in living cells such as cell locomotion. Moving on into this
direction will require a number of important additional ingredients. A three
dimensional description should, for example, incorporate effects, such as the
polymerization and depolymerization of filaments, non-mobile cross-linkers,
capping proteins, and the interaction of filaments with a membrane.

\begin{acknowledgments}
We thank J.~Prost, S.~Camalet, and K.~Sekimoto for stimulating discussions.
K.~K. acknowledges financial support by the Max-Planck-Gesellschaft through a
Schl\"o\ss mann fellowship as well as the kind hospitality of the Landau
Institute for Theoretical Physics, Moscow.
\end{acknowledgments}

\appendix

\section{\label{app:eigenwerte}Eigenvalues of the linearized
time-evolution operator}

Here, we give the complete eigenvalues of the linearized time-evolution
operator $\Lambda(k)$ of the minimal model given in Eq.~(\ref{eq:fmmmo}) and
show, that $\alpha_c>0$.

The two eigenvalues of $\Lambda(k)$ are
\begin{widetext}
\begin{eqnarray}
  \label{eq:eigenwerte} \lambda_\pm & = & -k^2 + \alpha(1-\cos k)c - 2i\beta k
  \delta c \nonumber\\ & & \pm\left\{ \alpha^2(1-\cos k)^2{\delta c}^2 -
  \beta^2 k^2c^2 +\beta^2\sin^2k (c^2-{\delta c}^2) + 2i\alpha\beta k(1-\cos
  k)c\delta c\right\}^{1/2}\quad.
\end{eqnarray}
\end{widetext}
In this expression $c=c^+_0+c^-_0$ and $\delta c=c^+_0-c^-_0$. The real part
of $\lambda_+$, which determines the stability of the homogeneous state
against small perturbations, is
\begin{equation}
  \label{eq:lambda} \lambda(k) = -k^2 + \alpha(1-\cos k)c +
  \left\{\frac{1}{2}\sqrt{a^2+b^2} + \frac{1}{2}a\right\}^{1/2}
\end{equation}
with
\begin{eqnarray}
  a & = & \alpha^2 (1-\cos k)^2 {\delta c}^2 \nonumber\\
&&- \beta^2 k^2 c^2 +
  \beta^2\sin^2k(c^2 - {\delta c}^2) \\ b & = & 2\alpha\beta k(1-\cos k)
  c\delta c\quad.
\end{eqnarray}
For $\alpha=0$ this implies $ \lambda(k) = -k^2 $.  The derivative of
$\lambda$ with respect to $\alpha$ is of the form
\begin{equation}
  \label{eq:lambdaabl} \frac{\partial \lambda}{\partial \alpha} = A_0 +
  A_1\alpha + A_2 \alpha^3
\end{equation}
with $A_i>0$, $i=0,1,2$.  It then follows that there is a unique critical
value $\alpha_c>0$, determined by $\lambda(k=2\pi/L; \alpha=\alpha_c)=0$, such
that the homogenous state is linearly stable unless $\alpha>\alpha_c$ and that
the longest wave-length fitting in the system becomes unstable.

\section{\label{app:stoerung}Non-homogeneous solutions close to
\protect{$\alpha_c$}}

We calculate the asymptotic solutions of the minimal model close to the
critical value  $\alpha_c$ for arbitrary $\beta$ and compare it to the
expressions for small $\beta$ discussed in sect. \ref{sec:model1}.

\subsection{\label{app:betaNull}Oriented bundles}

\noindent In Fourier-representation the stationary solution of
Eq.~(\ref{eq:partialC}) is given by 
\begin{equation}
c_k = -\frac{2\alpha}{k}\sum_p \frac{\cos p-1}{p}c_p c_{k-p}\quad,
\end{equation}
where $c_{-k}=c_k^*$. In particular
\begin{widetext}
\begin{eqnarray}
c_1 & = & -\frac{2\alpha L^2}{4\pi^2}\left\{
 \left[\cos\frac{2\pi}{L}-1\right]c_1 c_0  +
 \frac{1}{2}\left[\cos\frac{4\pi}{L}-1\right]c_2 c_{-1}  -
 \left[\cos\frac{2\pi}{L}-1\right]c_{-1} c_2  +\ldots\right\} \\ c_2 & = &
 -\frac{\alpha L^2}{4\pi^2}\left\{ \left[\cos\frac{2\pi}{L}-1\right]c_1 c_1  +
 \frac{1}{2}\left[\cos\frac{4\pi}{L}-1\right]c_2 c_{0}  +\ldots\right\}
\end{eqnarray}
\end{widetext}
Close to $\alpha=\alpha_c$, we expand the solution in terms of $c_1$ using the
{ansatz} $c_k\propto c_1^k$.  Ignoring higher modes, we obtain
\begin{equation}
c_2 = -\frac{1}{1+\frac{\alpha L^2}{4\pi^2}\frac{1}{2}\left[
\cos\frac{4\pi}{L}-1\right]c_0}\frac{\alpha L^2}{4\pi^2}
\left[\cos\frac{2\pi}{L}-1\right] c_1^2.
\end{equation}

\noindent Up to third order in $c_1$ we find\\
\begin{equation}
F(\alpha)c_1 -G(\alpha)|c_1|^2 c_1 = 0,
\end{equation}
with
\begin{equation}
\label{eq:Fbetanull}
F(\alpha) = 1 + \frac{2\alpha L^2}{4\pi^2} \left[\cos\frac{2\pi}{L}-1\right]
 c_0
\end{equation}
and
\begin{widetext}
\begin{equation}
\label{eq:Gbetanull}
G(\alpha) = \frac{2\alpha L^2}{4\pi^2}\left\{
 \frac{1}{2}\left[\cos\frac{4\pi}{L}-1\right]-\left[\cos\frac{2\pi}{L}-1\right]
 \right\} \frac{\frac{\alpha L^2}{4\pi^2}\left[\cos\frac{2\pi}{L}-1\right]}
 {1+\frac{\alpha L^2}{4\pi^2}\frac{1}{2}\left[ \cos\frac{4\pi}{L}-1\right]c_0}.
\end{equation}
%\end{widetext}
%\begin{widetext}
The Fourier coefficients of the stationary solution are given by
\begin{equation}
c_1 = \frac{L c_0}{2\pi}\sqrt{\frac{-2c_0}{\cos\frac{2\pi}{L}}}
 \left[1-\cos\frac{2\pi}{L}\right]\left(\alpha-\alpha_c\right)^{1/2} +
 O\left(\left(\alpha-\alpha_c\right)^{3/2}\right) \label{eq:C1}
\end{equation}
and
\begin{equation}
c_2=-\frac{2L^2c_0^2}{4\pi^2}\frac{1-\cos\frac{2\pi}{L}}{\cos\frac{2\pi}{L}}
 \left(\alpha-\alpha_c\right) +
 O\left(\left(\alpha-\alpha_c\right)^2\right).\label{eq:C2}
\end{equation}
\end{widetext}

\subsection{Bundles of mixed orientation}
\label{app:betaUnglNull}

We determine solitary waves which appear at the bifurcation point
$\alpha=\alpha_c$ using the ansatz
\begin{equation}
c^\pm(x,t) = \sum c_{k}^\pm {\rm e}^{ik(2\pi x/L+\omega t)}.
\end{equation}
The Fourier coefficients satisfy the  following equations:
\begin{widetext}
\begin{eqnarray}
\label{eq:c1Pl}
i\omega c_1^+ & = & -\frac{4\pi^2}{L^2}c_1^+
  +2\alpha\left\{\left[1-\cos\frac{2\pi}{L}\right]
  \left(c_1^+c_0^+-c_{-1}^+c_2^+\right)
  +\frac{1}{2}\left[1-\cos\frac{4\pi}{L}\right]c_{-1}^+c_2^++\ldots\right\}
  \nonumber\\ & & +2i\beta\left\{\frac{2\pi}{L}c_0^-c_1^+
  +\sin\frac{2\pi}{L}\left(c_1^-c_0^++c_{-1}^-c_2^+\right)
  +\frac{1}{2}\sin\frac{4\pi}{L}c_2^-c_{-1}^++\ldots\right\}\\ 2i\omega c_2^+
  & = & -\frac{16\pi^2}{L^2}c_2^+
  +4\alpha\left\{\left[1-\cos\frac{2\pi}{L}\right]c_1^+c_1^+
  +\frac{1}{2}\left[1-\cos\frac{4\pi}{L}\right]c_2^+c_0^++\ldots\right\}
  \nonumber\\ & & +4i\beta\left\{\frac{2\pi}{L}c_0^-c_2^+
  +\sin\frac{2\pi}{L}c_1^-c_1^+
  +\frac{1}{2}\sin\frac{4\pi}{L}c_2^-c_{0}^++\ldots\right\}\\ i\omega c_1^- &
  = & -\frac{4\pi^2}{L^2}c_1^-
  +2\alpha\left\{\left[1-\cos\frac{2\pi}{L}\right]
  \left(c_1^-c_0^--c_{-1}^-c_2^-\right)
  +\frac{1}{2}\left[1-\cos\frac{4\pi}{L}\right]c_{-1}^-c_2^-+\ldots\right\}
  \nonumber\\ & & -2i\beta\left\{\frac{2\pi}{L}c_0^+c_1^-
  +\sin\frac{2\pi}{L}\left(c_1^+c_0^-+c_{-1}^+c_2^-\right)
  +\frac{1}{2}\sin\frac{4\pi}{L}c_2^+c_{-1}^-+\ldots\right\}\\ 2i\omega c_2^-
  & = & -\frac{16\pi^2}{L^2}c_2^-
  +4\alpha\left\{\left[1-\cos\frac{2\pi}{L}\right]c_1^-c_1^-
  +\frac{1}{2}\left[1-\cos\frac{4\pi}{L}\right]c_2^-c_0^-+\ldots\right\}
  \nonumber\\ & & -4i\beta\left\{\frac{2\pi}{L}c_0^+c_2^-
  +\sin\frac{2\pi}{L}c_1^+c_1^-
  +\frac{1}{2}\sin\frac{4\pi}{L}c_2^+c_{0}^-\ldots\right\}.
\label{eq:c2Mi}\\
& \vdots &\nonumber
\end{eqnarray}
\end{widetext}

We assume without loss of generality $c_0^+>c_0^-$ and expand in  $c_1^+$,
which leads to
\begin{equation}
F(\alpha, \beta, \omega) - G(\alpha, \beta, \omega)|c_1^+|^2 = 0,
\end{equation}
where the expressions for $F$ and $G$ to first order in $\beta$ are given by
\begin{widetext}
\begin{equation}
F = -\frac{4\pi^2}{L^2}+2\alpha\left[1-\cos\frac{2\pi}{L}\right]c_0^+
    +i\left[\frac{4\pi}{L}\beta c_0^--\omega\right]
\end{equation}
and
\begin{equation}
G = 8\alpha^2\frac{L^2}{4\pi^2}\left[1-\cos\frac{2\pi}{L}\right]
\left\{\left[1-\cos\frac{2\pi}{L}\right] -
\frac{1}{2}\left[1-\cos\frac{4\pi}{L}\right]\right\}
\left\{\frac{1}{\Delta}-2i\frac{\omega_1-\frac{4\pi^2}{L}c_0^-}{\Delta^2}\beta\right\}
\end{equation}
\end{widetext}
Here,  $\Delta=16\pi^2/L^2 - 2\alpha\left[1-\cos\frac{4\pi}{L}\right]c_0^+$.

At the bifurcation point $\alpha=\alpha_c$ we have
$F(\alpha_c,\beta,\omega_0)=0$. In first order in $\alpha-\alpha_c$ we obtain
for the frequency and the amplitude of $c_1^+$
\begin{eqnarray}
\omega & = & \omega_0 + \omega_1(\alpha-\alpha_c) + \ldots\\ |c_1^+|^2 & = &
\frac{\partial_\alpha F(\alpha_c,\omega_0) +  \omega_1\partial_\omega
F(\alpha_c,\omega_0)}{G(\alpha_c,\omega_0)} (\alpha-\alpha_c) + \ldots
\end{eqnarray}
The right hand side of the last expression needs to be real and positive such
that we obtain the condition
\begin{equation}
\textrm{Im} \frac{\partial_\alpha F(\alpha_c,\omega_0)}{G(\alpha_c,\omega_0)}
= \omega_1 \textrm{Im} \frac{\partial_\omega
F(\alpha_c,\omega_0)}{G(\alpha_c,\omega_0)} .
\end{equation}
Furthermore, the sign of the prefactor of $\alpha-\alpha_c$ will determine if
the bifurcation is super- oder sub-critical.

The bifurcation condition $F(\alpha_c,\omega_0) = 0$ leads to
\begin{eqnarray}
\alpha_c & = &
\frac{4\pi}{L^2}\frac{1}{2\left[1-\cos\frac{2\pi}{L}\right]c_0^+} \\  \omega_0
& = & \frac{4\pi}{L}\beta c_0^-\quad .
\end{eqnarray}
This implies
\begin{eqnarray}
\partial_\alpha F(\alpha_c,\omega_0) & = & \frac{2L^2}{4\pi^2}\left[
1-\cos{2\pi}{L}\right]c_0^+ \\ \partial_\omega F(\alpha_c,\omega_0) & = &
-i\frac{L^2}{4\pi^2}\\
\label{eq:Gacomnull}
G(\alpha_c,\omega_0) & = &
-\frac{1}{{c_0^2}^2}\frac{\cos\frac{2\pi}{L}}{1-\cos\frac{2\pi}{L}} \quad ,
\end{eqnarray}
and $\omega_1=0$.

The velocity of propagation is given by $v=\omega_0 L/2\pi=2\beta c_0^-$. The
components $c_1^+$ and $c_2^+$ are given by the expressions (\ref{eq:C1}) and
(\ref{eq:C2}) for $c_1$ and $c_2$, respectively.  For the distribution of
minus-filaments we find
\begin{widetext}
\begin{eqnarray}
c_1^- & = & -2i\beta\frac{\sin\frac{2\pi}{L}c_0^-}
{\frac{4\pi^2}{L^2}-2\alpha\left[1-\cos\frac{2\pi}{L}\right]c_0^-}c^+_1
\label{eq:C1minus}\\
c_2^- & = & i\beta\frac{\sin\frac{2\pi}{L}\left[1-\cos\frac{2\pi}{L}\right]^2
{c_0^+}^2c_0^-}{\frac{16\pi^4}{L^4}\left\{1
-\frac{1}{2}\left[1+\cos\frac{2\pi}{L}\right]\frac{c_0^++c_0^-}{c_0^+}+
\frac{1}{4}\left[1+\cos\frac{2\pi}{L}\right]^2\frac{c_0^-}{c_0^+}\right\}}
(\alpha-\alpha_c) + O\left((\alpha-\alpha_c)^2\right).
\label{eq:C2minus}
\end{eqnarray}
\end{widetext}

These expressions are consistent with the results of
Sec.~\ref{sec:betaUnglNull} and provide analytic expressions of solitary wave
solutions for small $\beta$.

\section{\label{sec:adhaesion}Filament-Adhesion} 

In this section, we discuss the effects on the bundle dynamics due to filament
adhesion to a substrate in the minimal model.  Adhering filaments are
described by the densities $a^+$ and $a^-$, respectively.  Attached filaments
are assumed to be immobile, while they contribute to the motion of free
filaments.  The dynamic equations in dimensionless form are given by
\begin{eqnarray}
\partial_t c^+ & = & \partial_x^2 c^+ -\partial_x J^{++} - \partial_x J^{+-}
-\omega_a c^+ +\omega_d a^+ \label{eq:adh1} \\  \partial_t c^- & = &
\partial_x^2 c^- -\partial_x J^{--} - \partial_x J^{-+} -\omega_a c^-
+\omega_d a^- \\  \partial_t a^+ & = & \omega_a c^+ -\omega_d a^+ \\
\partial_t a^- & = & \omega_a c^- -\omega_d a^- \label{eq:adh4}
\end{eqnarray}
with
\begin{widetext}
\begin{equation}
J^{\pm\pm}  =  \int_0^1 d\xi\; \left[\alpha c^\pm(x+\xi)+\bar\alpha
a^\pm(x+\xi)  -\alpha c^\pm(x-\xi)-\bar\alpha a^\pm(x-\xi)\right]c^\pm(x)
\end{equation}
and
\begin{equation}
J^{\pm\mp}  = \mp\int_{-1}^1 d\xi\; \left[\beta c^\mp(x+\xi)+\bar\beta
  a^\mp(x+\xi)\right] c^\pm(x) \quad.
\end{equation}
\end{widetext}
Here $\omega_a$ and $\omega_d$ are rates of attachment and detachment of
filaments, respectively while $\bar\alpha$ and $\bar\beta$  characterize
coupling constants between free and attached filaments.

\begin{figure}[t]
  \includegraphics[width=\linewidth]{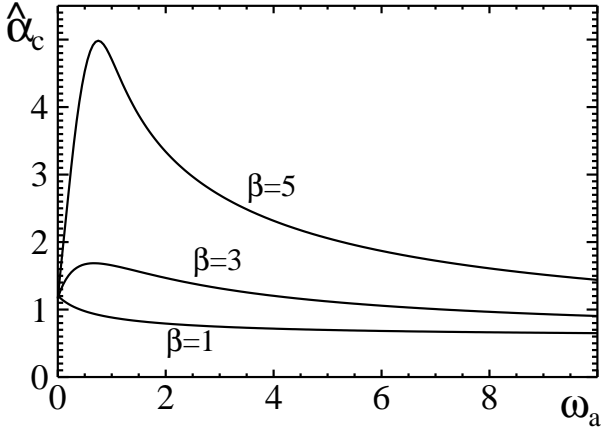} \caption{The effect of filament
  adhesion on the 
  critical value $\hat\alpha_c$ of the minimal model. Displayed is the
  critical value $\hat\alpha_c$ as a function of the adhesion rate $\omega_a$
  for different values of the coupling strength $\beta$ and for $c_0^+=0.3$,
  $c_0^-=0.7$, $\bar\alpha=2\alpha$, $\bar\beta=2\beta$, $\omega_d=1$, and
  $L=10$. For $\omega_a=0$ $\hat\alpha_c=\alpha_c$, which is the critical
  value in the minimal model in the absence of adhesion.}
  \label{fig:linstabAdh}
\end{figure}

The homogeneous state $c^\pm(x)=c^\pm_0$ and
$a^\pm(x)=\omega_ac^\pm_0/\omega_d$ is stationary. It becomes unstable at a
critical value $\alpha_c$ which for oriented bundles is given by
\begin{equation}
\hat\alpha_c = \alpha_c\frac{\omega_a+\omega_d}{\mu\omega_a+\omega_d},
\end{equation}
where $\alpha_c$ is the critical value in the minimal model with $\omega_a=0$
and for $\mu=\bar\alpha/\alpha$.  For $\beta\neq0$ a Hopf-bifurcation occurs,
the  dependence of the critical value  $\hat\alpha_c$ on the attachment rate
$\omega_a$ is shown in Fig.~\ref{fig:linstabAdh} for $\bar\alpha=2\alpha$ and
$\bar\beta=2\beta$.

\begin{figure}[t]
  \includegraphics[width=\linewidth]{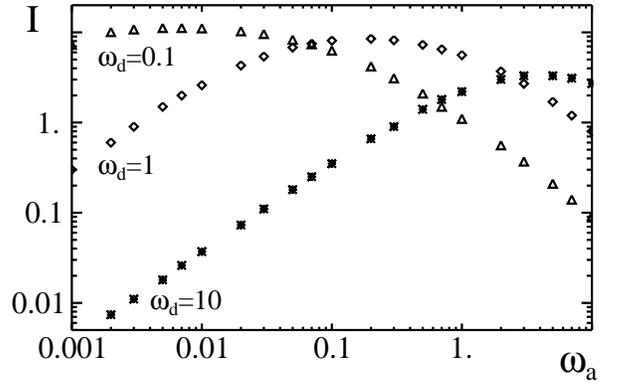} \caption{The total filament
  current $I$ of 
  propagating filament patterns with filament adhesion as a function of the
  adhesion rate $\omega_a$. Symbols represent numerically obtained results for
  different detachment rates $\omega_d$. Parameters are $\alpha=1.5$,
  $\beta=2$, $\bar\beta=2\beta$, $c_0^+=0.3$, $c_0^-=0.7$, and $L=10$. The
  maximal current decreases and is reached at larger ratios of
  $\omega_a/\omega_d$ as $\omega_d$ is increased. } \label{fig:strom}
\end{figure}
In the case $\beta=0$, the stationary distributions are given by
$c^\pm(x;\omega_a)=\frac{\omega_d}{\mu\omega_a+\omega_d} c^\pm(x;0)$ and
$a^\pm(x;\omega_a)=\frac{\omega_a}{\mu\omega_a+\omega_d}  c^\pm(x;0)$ where
$c^\pm(x;0)$ denote the stationary states of the minimal model.  For
$\beta\neq0$ solutions can be obtained in the limit of small attachment rates
by expanding around solitary waves for $\omega_a=0$. Consider a solitary wave
in the minimal model given by $c^\pm(x,t;0)=u_0^\pm(x-v_0t)$ and
$a^\pm(x,t;0)=0$. Assuming that for finite $\omega_a$ solutions of the form
$c^\pm(x,t;\omega_a)=u^\pm(x-vt)$ and $a^\pm(x,t:\omega_a) =r^\pm(x-vt)$
exist, we write
\begin{eqnarray}
  v & = & v_0 + \omega_a v_1 + \ldots\\ \label{eq:entwicklungAdh1} u^\pm & = &
  u^\pm_0 + \omega_a u^\pm_1 +\ldots\\ \label{eq:entwicklungAdh2} r^\pm & = &
  \omega_a r_1^\pm +\ldots
\end{eqnarray}
The equations for $r^\pm_1$ can be solved explicitly. We find
\begin{equation}
  \label{eq:asymp} r^\pm_1(x) =
  \frac{1}{\textrm{e}^{\omega_dT}-1}\int_0^Tdt^\prime\; {\rm
  e}^{\omega_dt^\prime} u^\pm_0(x-v_0t^\prime)\quad,
\end{equation}
where $T=v_0L$.

This result can in turn be used to calculate in first order in $\omega_a$ the
total net current $I$ associated with these solitary waves. Indeed, in lowest
order this current is given by
\begin{widetext}
\begin{eqnarray}
  \label{eq:stromtotal} I & = & 2\omega_a\!\!\int_0^L
  \!\!dx\left\{\alpha\!\!\int_0^1 \!\!d\xi
  \left[\left(r^+_1(x+\xi)-r^+_1(x-\xi)\right)u^+_0(x) +
  \left(r^-_1(x+\xi)-r^-_1(x-\xi)\right)u^-_0(x)\right]\right.\nonumber\\ & &
  \quad\quad\quad\quad\quad +
  \beta\int_{-1}^1\!d\xi\left.\left[r^+_1(x+\xi)u^-_0(x) -
  r^-_1(x+\xi)u^+_0(x)\right]\right\}\quad,
\end{eqnarray}
\end{widetext}
where, for simplicity, we have chosen $\bar\alpha=2\alpha$ and
$\bar\beta=2\beta$. Evaluating this expression using Eq.~(\ref{eq:asymp}) and
$u_0^\pm(x)=\sum_{n=-\infty}^\infty u_{0,n}^\pm{\rm e}^{i2\pi nx/L}$ we obtain
\begin{equation}
  \label{eq:stromerstord} I = 8\omega_a L^3v_0\alpha\sum_{n=1}^\infty
  A_n\left(|u^+_{0,n}|^2+|u^-_{0,n}|^2\right) + O(\omega_a^2)\quad,
\end{equation}
where
\begin{equation}
  \label{eq:An} A_n = \frac{\cos\frac{2\pi n}{L}-1}{\omega_d^2L^2 +
  4\pi^2n^2v_0^2}
\end{equation}
with $v_0$ and $u_0^\pm$ depending on $\beta$ as described in
Sec.~\ref{sec:betaUnglNull}.  This result shows, that  solitary waves are
accompanied with a net filament transport if filament adhesion occurs. Since
$A_n<0$ for all $n$, this transport occurs into the opposite direction as wave
propagation. Remarkably, as indicated by the prefactor $\alpha$, it is the
interaction between {\em parallel} filaments that generates the current. Note,
that for $\beta=0$ we have $v_0=0$ and thus $I=0$.  Numerical results for $I$
as a function of $\omega_a$ are shown in Fig.~\ref{fig:strom}.

\end{document}